# Origin of the fast magnetization tunneling in the single-molecule magnet [Ni(hmp)(tBuEtOH)Cl]$_4$

C. Kirman,[a] J. Lawrence,[a] S. Hill,[a] E-C. Yang,[b] and D. N. Hendrickson[b]

[a]Department of Physics, University of Florida, Gainesville, FL 32611
[b]Department of Chemistry and Biochemistry, University of California at San Diego, La Jolla, CA 92093

We present high-frequency angle-dependent EPR data for crystals of [Ni$_x$Zn$_{1-x}$(hmp)(t-BuEtOH)Cl]$_4$ ($x = 1$ and 0.02). The $x = 1$ complex behaves as a single-molecule magnet at low temperatures, displaying hysteresis and exceptionally fast magnetization tunneling. We show that this behavior is related to a 4th-order transverse crystal-field interaction, which produces a significant tunnel-splitting (~10 MHz) of the ground state of this $S = 4$ system. The magnitude of the 4th-order anisotropy, and the dominant axial term ($D$), can be related to the single-ion interactions ($D_i$ and $E_i$) at the individual Ni$^{II}$ sites, as determined for the $x = 0.02$ crystals.

Single-molecule magnets (SMMs) provide a molecular, or "bottom-up," route to magnetic nanostructures [1]. They typically consist of a core of magnetically coupled transition metal ions, resulting in a well-defined giant spin ground state. Their main attraction is an intrinsic bistability, which is realized via a significant negative (easy-axis) magnetocrystalline anisotropy [see Eq. 1]. This bistability has aroused significant interest in terms of the use of SMMs in future molecular devices [2]. Unlike mesoscale magnetic particles of much larger dimensions, SMMs straddle the interface between classical and quantum behavior, displaying both superparamagnetic-like behavior, and magnetic quantum tunneling (MQT) at low temperatures. Furthermore, they offer all of the advantages of molecular chemistry, e.g. purity, solubility in various solvents, a well defined periphery of organic ligands, a crystalline assembly of monodisperse units, etc.. When grown as crystals, the magnetic unit is monodisperse—each molecule in the crystal has the same spin, orientation, magnetic anisotropy and structure, etc. [1]. Thus, bulk probes of SMM crystals enable fundamental studies of properties intrinsic to individual magnetic nanostructures.

We have recently synthesized a family of SMMs based on tetranuclear nickel clusters having the general formula [Ni(hmp)($R$OH)X]$_4$, where $R$ = CH$_3$, C$_2$H$_5$, etc., and X = Cl or Br [3], and hmp$^-$ is the monoanion of 2-hydroxymethylpyridine. Low-temperature ($\ll$ 1 K) magnetization studies have demonstrated that each of these Ni$_4$ complexes exhibit MQT [3], albeit the temperature-independent MQT rate is exceptionally fast (too fast to measure). In this article, we focus on the [Ni(hmp)(t-BuEtOH)Cl]$_4$ member of this family, where t-BuEtOH is 3,3-dimethyl-1-butanol. Previous electron paramagnetic resonance (EPR) studies have confirmed the expected $S$ = 4 ground state, as well as showing that it possesses an easy-axis-type anisotropy [3,4], i.e. a negative axial crystal-field parameter ($D \approx$ -0.600 cm$^{-1}$). Our motivation for studying the $R$ = t-BuEtOH complex stems partly from its high symmetry ($S_4$). More importantly, however, crystals of this complex contain no solvate molecules. As recent studies of Mn$_{12}$-acetate have shown, solvent molecules can have a pronounced influence on the quantum properties of SMMs [5,6]. The Ni$_4$ SMMs offer no exception. EPR spectra for the $R$ = CH$_3$ and C$_2$H$_5$ Ni$_4$ complexes consist of extremely broad absorption peaks, with multiple fine structures, making detailed analysis rather difficult [4]. In contrast, EPR spectra for the t-BuEtOH complex exhibit very sharp lines (see Fig. 1 and [4]).

The effective spin Hamiltonian for an isolated Ni$_4$ SMM is [1]:

$$\hat{H} = D\hat{S}_z^2 + g\mu_B \bm{B} \cdot \hat{S} + \hat{O}_4 + \hat{H}'; \qquad (1)$$

where $\hat{S}$ is the spin angular momentum operator with components $\hat{S}_x$, $\hat{S}_y$, and $\hat{S}_z$; $g$ is the Landé g-factor and $\bm{B}$ is the applied field strength; $\hat{O}_4$ denotes fourth-order crystal-field terms; and $\hat{H}'$ describes additional perturbations such as hyperfine and intermolecular interactions [1-5]. MQT is caused by terms in the Hamiltonian which do not commute with the dominant $D\hat{S}_z^2$ interaction, e.g. terms containing $\hat{S}_x$ and $\hat{S}_y$. The rhombic term, $E(\hat{S}_x^2 - \hat{S}_y^2)$, is symmetry forbidden. Thus, the leading term in the crystal-field that could cause MQT would be the fourth-order $B_4^4(\hat{S}_+^4 - \hat{S}_-^4)$ interaction. We note that this operator connects states differing in $m_s$ (projection of $S$ along $z$) by ±4 in second order of perturbation theory. Consequently, it is quite effective at causing tunneling between the $m_s = \pm 4$ ground states. While its effects are rather weak at low fields, causing tunnel splittings of order 10 MHz (see below), the interaction becomes zeroth-order in the presence of a large transverse field ($g\mu_B B > DS$). In this way, angle-dependent hard-plane EPR experiments can be used to determine both the symmetry of the dominant transverse terms in Eq. 1 [5,6], and to quantify their magnitudes (note: $D$ and $g$ are isotropic in the $xy$-plane). Similarly, angle-dependent EPR studies of a doped [Ni$_x$Zn$_{1-x}$(hmp)(t-BuEtOH)Cl]$_4$ ($x$ = 0.02) crystal enable a determination of the complete crystal-field tensor associated with each Ni$^{II}$ site ($D_i$ and $E_i$ parameters, and associated Euler angles), which one can then relate to the full Ni$_4$ spin Hamiltonian [7].

High-frequency (40–300 GHz) single-crystal EPR measurements were carried out using a millimeter-wave vector network analyzer (MVNA) in combination with an over-moded cavity perturbation technique which we have described elsewhere [8]. In order to enable in-situ rotation of the sample relative to the applied magnetic field, we employed a split-pair magnet with a 7 T horizontal field and a vertical access. Smooth rotation of the entire rigid microwave probe, relative to the fixed field, was achieved via a room temperature stepper motor. Synthetic procedures for obtaining single crystals of [Ni$_x$Zn$_{1-x}$(hmp)(t-BuEtOH)Cl]$_4$ ($x$ = 1 and 0.02) have been described previously [3]. For the angle-dependent measurements, pyramidal-shaped single-crystal samples [approximate dimensions: 1.5 × 1.5 mm$^2$ square base; 2 mm height (see insets to figures)] were aligned by hand within a cylindrical cavity (dia. = 7.62 mm, length = 8.38 mm) for field rotations in the appropriate crystallographic plane. Small mis-alignments are easily corrected during data analysis (see below). In all cases, the temperature was stabilized relative to a calibrated Cernox$^{TM}$ resistance sensor using a combination of heaters and cold helium gas flow.

Figure 1(a) displays the microwave transmission through the cavity at a temperature of 10 K and a frequency of 101.2 GHz, for the $x$ = 1 complex; in this experiment, the magnetic field was rotated in the hard [$xy$ or (100)/(010)] plane of the crystal. The data in Fig. 1(a) correspond to the case where the field is applied 23$^o$ away from one of the edges of the square base of the pyramidal sample [$x$ or $y$ direction, see inset to (b)], corresponding to the hard axis of the crystal ($\phi$ = 0). The sharp minima in transmission correspond to EPR absorptions; these resonances have been labeled according to the scheme described in Ref. [6]. The splitting of the highest field peak (also weakly visible in the 4.5 T peak) is caused by a weak disorder associated with the t-BuEtOH ligand which gives rise to micro-environments with slightly different $D$ values (-0.600 cm$^{-1}$ and -0.577 cm$^{-1}$ [4]). This disorder, which sets in below a structural transition at 46 K, has been confirmed via specific heat and X-ray measurements [9].

Figure 1(b) displays a grayscale contour plot of the absorption intensity, as a function of the magnetic field strength and its orientation ($\phi$) within the hard plane. Immediately apparent is the 4-fold behavior of the peak position shifts. The maxima occur when the field is along the hard directions ($\phi = 0°$, 90°, etc.). The minima correspond the medium axes, which are in between the hard directions ($\phi = 45°$, 135°, etc.). The hard directions ($\phi = 0°$, 90°, etc.) are located at -23° (or +23°) and 67° (or -67°) away from the crystallographic (100) and (010) directions. The 4-fold line-shifts are caused by the fourth-order $B_4^4(\hat{S}_+^4 - \hat{S}_-^4)$ interaction in Eq. (1). Superimposed on the absorption maxima (darker regions) in Fig. 1(b) are fits (white curves) to the data. The 4-fold shifts are reproduced for all peaks (including β4´) with just a single parameter, $B_4^4 = 4 \times 10^{-4}$ cm$^{-1}$. In addition, the hard-plane g-factors ($g_x = g_y = 2.23$) may be obtained from the average peak positions. The satellite peak (β4´) between 3.15 T and 3.3 T, which is only seen for angles close to $\phi = 0°$, is a double quantum transition [all others in Fig. 1b are $\Delta m_s = 1$]. This transition is only allowed when there is a significant microwave H$_1$-field component parallel to the applied DC field H$_0$. It turns out that the relative orientations of H$_1$ and H$_0$ vary upon rotation of the cavity such that they are approximately parallel for $\phi = 0°$ and 180°, and perpendicular for $\phi = 90°$ and 270° [10]. Spectra obtained at different $\phi$ angles were normalized to the intensity of α4, hence β4´ vanishes in the vicinity of $\phi = 90°$, i.e. the 2-fold nature of this peak is an instrumental artifact and not connected to any intrinsic 2-fold symmetry of the sample. Finally, we note that the obtained large fourth-order $B_4^4$ term results in a 10 MHz tunnel splitting between the $m_s = \pm 4$ ground states, thus explaining the fast magnetization tunneling in this system.

Figure 2(a) displays a series of EPR spectra for a doped [Ni$_{0.02}$Zn$_{0.98}$(hmp)(tBuEtOH)Cl]$_4$ crystal at 2.5 K and 120 GHz. The field was rotated in the $xz$ plane: in the top trace, the field is oriented at $\theta = -34°$ away from the hard plane; successive traces were taken in 2.5° steps towards the hard plane. Fig. 2(b) displays the angle-dependence of the EPR peak positions for a full 360° rotation of the sample (inset depicts the experimental geometry). Due to the low concentration of Ni$^{II}$ in the sample, we can assume that the EPR spectra are dominated by NiZn$_3$(hmp)$_4$(t-BuEtOH)$_4$Cl$_4$ molecules having $S = 1$. However, the Ni$^{II}$ ion can reside at four different sites on the molecule and, in general, the crystal-field tensor at each site will have a different orientation, albeit related by the symmetry group of the molecule. Indeed, this is the reason why one observes a pair of maxima either side of $\theta = 0°$ and 180° in Fig. 2(b); these maxima would coincide at $\theta = 0°$ and 180° if the Ni$^{II}$ crystal-field tensors were collinear and coincident with the full Ni$_4$ crystal-field tensor. Thus, these experiments indicate that the Ni$^{II}$ tensors are tilted with respect to the molecule. In fact, all of the peaks in Fig. 2 correspond to the same transition, as indicated in the $S = 1$ energy level diagram (inset to Fig. 2a). The four resonance branches originate from each of the four inequivalent Ni$^{II}$ sites. For the ideal experimental geometry, one should expect the two high-field branches to lie on top of each other. However, a slight sample mis-alignment lifts this degeneracy. Nevertheless, this is taken into consideration in our fitting procedure. Similar experiments performed in a plane parallel to the diagonals of the pyramidal sample (inclined 45° to the data in Fig. 2), reveal that the Ni$^{II}$ easy axes are tilted along the (100) and (010) directions, and inclined 15° away from the (001) direction. This likely explains why the hard directions for the Ni$_4$ molecule are found between these directions (see Fig. 1), i.e. along (100) and (010).

Frequency-dependent studies (not shown) performed with the field along one of the high symmetry directions provide access to the crystal-field parameters for the Ni$^{II}$ ions: $D_i = -5.27$ cm$^{-1}$ and $E_i = -1.2$ cm$^{-1}$. A simple calculation, which assumes no tilting, enables a comparison between $D_i$ and the Ni$_4$ $D$ value: the negative E value yields $D = -0.66$ cm$^{-1}$; the positive E value yields $D = -0.69$ cm$^{-1}$ [7]. Such agreement (within 10%) is impressive, considering the approximate nature of the calculation. Armed with Ni$^{II}$ crystal field parameters, one can also fit the angle-dependent data for NiZn$_3$(hmp)$_4$(t-BuEtOH)$_4$Cl$_4$ for each plane of rotation [i.e. solid curves in Fig. 2(b)]. As discussed above, our fitting procedure allows for mis-alignments between the actual and assumed rotation planes; indeed, such a mis-alignment of about 8$^o$ is the reason for the asymmetry between the heights of the maxima in Fig. 2(b). A full account of this analysis will be presented elsewhere [7]. However, the easy axis tilting and significant $E$ term associated with the Ni$^{II}$ ions provide a natural explanation for the significant $B_4^4$ term and the fast magnetization relaxation for the Ni$_4$ system.

This work was supported by the National Science Foundation (DMR0103290 and DMR0239481). S.H. is a Cottrell scholar of the Research Corporation.

**Figure captions**

Fig. 1(a) Transmission through the cavity containing a [Ni (hmp)(t-BuEtOH)Cl]$_4$ single-crystal at 10 K and 101.2 GHz; the field is applied 23° away from one of the edges of the square base of the pyramidal sample [*x* or *y* direction, see inset to (b)]. The resonances have been labeled according to the scheme described in ref. [6]; the primed peaks (e.g. β4´) represent double quantum transitions; and β4´´ originates from the downward sloping branch of β4 in the frequency versus field plot, i.e. β4 is seen at two field positions (see Fig. 1 of ref. [6]). (b) Grayscale contour plot of the absorption intensity in (a), as a function of the magnetic field strength and its orientation ($\phi$) within the hard plane; the white curves are fits to the data (see main text for details).

Fig. 2(a) EPR spectra for a doped [Ni$_{0.02}$Zn$_{0.98}$(hmp)(t-BuEtOH)Cl]$_4$ crystal at 2.5 K and 120 GHz. The field was rotated in the *xz* plane: in the top trace, the field is oriented at $\theta = -34°$ away from the hard plane; successive traces were taken in 2.5° steps towards the hard plane. The inset shows the energy versus magnetic field (Zeeman) diagram for $S = 1$, for a field tilted 15° away from the easy axis; the levels are labeled in zero magnetic field ($|\pm1\rangle_S$ and $|\pm1\rangle_A$ correspond to symmetric and antisymmetric combinations of the $S = \pm1$ states) and the arrow indicates the origin of the EPR transitions seen in the main figure. (b) Angle-dependence of the EPR peak positions in (a) for a full 360° rotation of the sample (inset depicts the experimental geometry); the solid curves are fits to the data.

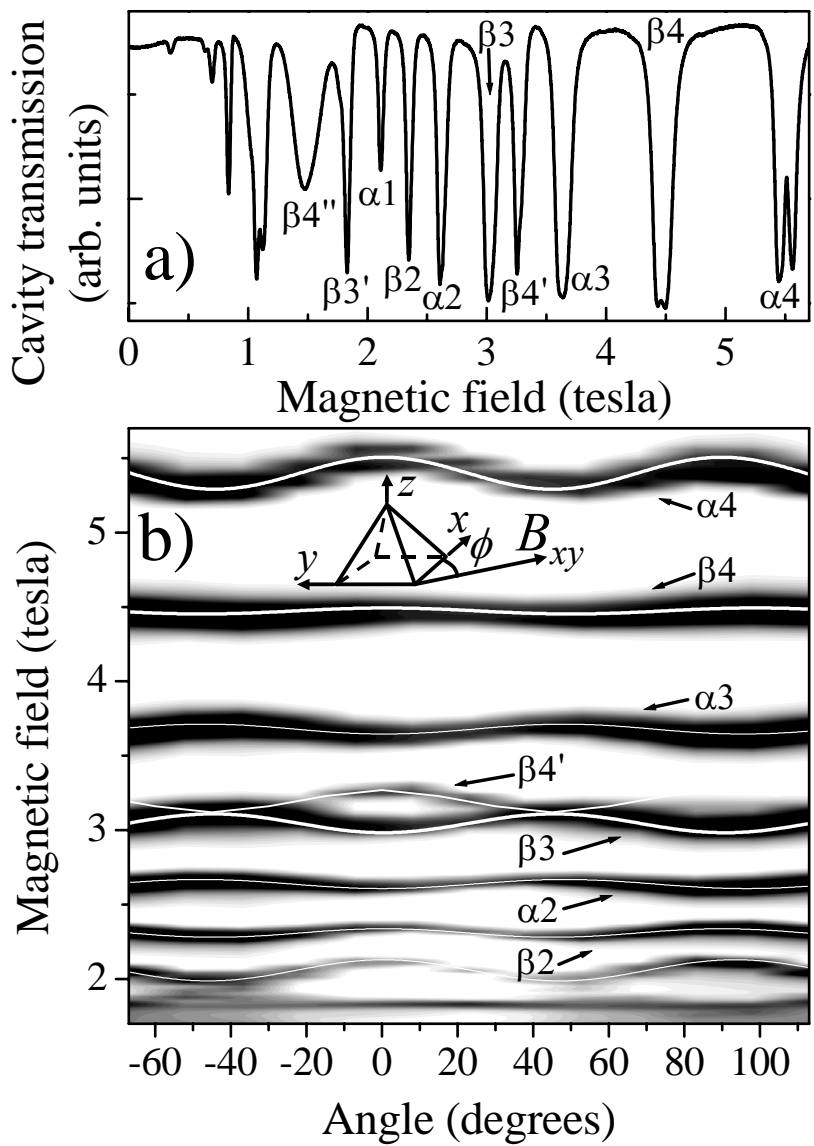

Kirman et al., Fig. 1

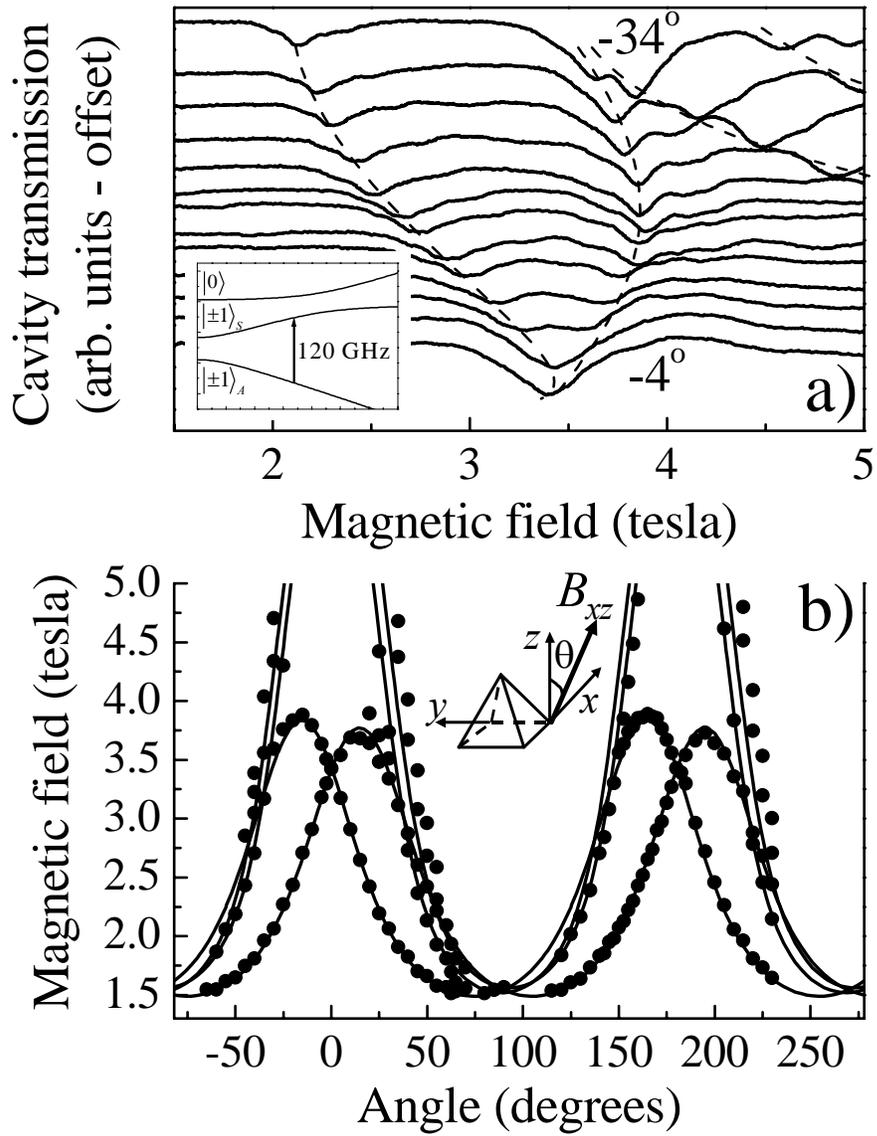

Kirman et al., Fig. 2